\documentclass[conference]{IEEEtran}
\usepackage{graphicx}
\usepackage{subfigure}
\usepackage{url}
\usepackage[noend, linesnumbered, boxed, ruled]{algorithm2e}
\usepackage{cases}
\usepackage{paralist}
\usepackage{balance}  
\usepackage{color}
\usepackage{listings}

\usepackage{amsthm}
\newcommand{\liststy}{\small\ttfamily}

\begin{document}


\title{Tolerating Correlated Failures in Massively Parallel Stream Processing Engines}

\author{\IEEEauthorblockN{Li Su} 
\IEEEauthorblockA{University of Southern Denmark\\
Email: lsu@imada.sdu.dk}
\and
\IEEEauthorblockN{Yongluan Zhou}
\IEEEauthorblockA{University of Southern Denmark\\
Email: zhou@imada.sdu.dk}
}

\maketitle

\begin{abstract}
Fault-tolerance techniques for stream processing engines can be categorized
into passive and active approaches. A typical passive approach periodically
checkpoints a processing task's runtime states and can recover
a failed task by restoring its runtime state using its latest checkpoint.  On the
other hand, an active approach usually employs backup nodes to run replicated
tasks. Upon failure, the active replica can take over the processing of the failed
task with minimal latency.  However, both approaches have their own
inadequacies in Massively Parallel Stream Processing Engines (MPSPE).  The
passive approach incurs a long recovery latency especially when a number of
correlated nodes fail simultaneously, while the active approach requires extra
replication resources. In this paper, we propose a new fault-tolerance
framework, which is Passive and Partially Active (PPA). In a PPA scheme, the passive
approach is applied to all tasks while only a selected set of tasks will be
actively replicated. The number of actively replicated tasks depends on the
available resources. If tasks without active replicas fail, tentative outputs
will be generated before the completion of the recovery process. We also propose
effective and efficient algorithms to optimize a partially active replication
plan to maximize the quality of tentative outputs. We implemented PPA on top of
Storm, an open-source MPSPE and conducted extensive experiments using
both real and synthetic datasets to verify the effectiveness of our approach.
\end{abstract}

\section{Introduction}

There is a recently emerging interest in building Massively Parallel
Stream Processing Engines (MPSPE), such as Storm
\cite{Storm:2014}, and Spark Streaming\cite{Zaharia:2013}, which make use
of large-scale computing clusters to process continuous queries over fast
data streams. Such continuous queries often run for a very long 
time and would unavoidably experience various system
failures, especially in a large-scale cluster. As it is critical to
provide continuous query results without significant downtime in many
data stream applications, fault-tolerance techniques in Stream Processing
Engines (SPEs)
\cite{Balazinska:2008,CastroFernandez:2013,Zaharia:2013} have
attracted a lot of attention.

Existing fault-tolerance techniques for SPEs can be generally
categorized as passive and active approaches~\cite{Hwang:2005}. In a
typical passive approach, the runtime states of tasks will be periodically
extracted as checkpoints and stored
at different locations. Upon failure, the state of a failed task can be
restored from its latest checkpoint. While one can in general tune
the checkpoint frequency to achieve trade-offs between the cost of
checkpoint and the recovery latency, the checkpoint frequency
should be limited to avoid high checkpoint overhead, which affects
the system performance. Hence recovery latency is usually significant
in a passive approach. 
When one wants to minimize the recovery latency as much as possible,
it is often more efficient to use an active approach, which typically uses one
backup node to replicate the tasks running on each processing node.  When a
node fails, its backup node can quickly take over with minimal latency. 

Even though there are abundant fault-tolerance techniques in SPEs, developing an
MPSPE \cite{Storm:2014} poses great challenges to the problem.
First of all, in a large cluster, there are often two different types of
failures: independent failure and correlated failure~\cite{Heath:2002,
Nath:2006}. Previous studies mostly focused on independent failure that
happens at a single node. 
Correlated failures are usually caused by failures of switches, routers and
power facilities, and will involve a number of nodes failing 
simultaneously. With such failures, one has to recover a large number of failed
tasks and temporarily run them on an additional set of standby nodes before the
failed ones are recovered. Using a passive fault tolerance approach, one has to
keep the standby nodes running even their utilization is low most of the time in
order to avoid the unacceptable overhead of starting them at recovery time.
Furthermore, as checkpoints of different nodes are often created asynchronously,
massive synchronizations have to be performed during recovery.
Therefore it could be difficult to meet the user requirements on recovery
latency even with a relatively high checkpoint frequency.

On the other hand, while an active fault-tolerance approach can achieve a lower
recovery latency, it could be too costly for a large-scale computation. Consider
a large-scale stream computation that is parallelized onto $100$ nodes, one may
not be able to afford another $100$ backup nodes for active replication.

Another challenge is that there exist some time-critical applications which 
prefer query outputs being generated in good time even if the outputs are
computed based on incomplete inputs. This kind of applications usually require
continuous query output for real-time opportune decision-making or
visualization. Consider a community-based navigation service, which collects and
aggregates user-contributed traffic data in a real-time fashion and then
continuously provides navigation suggestions to the users.  Failure of some
processing nodes could result in losing some user-contributed data. The
system, while waiting for the failed nodes to recover, can continue to help
drivers plan their routes based on the incomplete inputs. Other examples of such
applications are like intrusion detections, online visualization of real-time
data streams etc. Alerts of events matching the intrusion attack patterns or 
infographics generated over incomplete inputs are still meaningful to the users
and should be generated without any major delay. Consider the long recovery
latency for a large-scale correlated failure, the lack of trade-offs between
recovery latency and result quality would not be able to fulfill the
requirements of these applications.

To address the aforementioned challenges, we propose a new
fault-tolerance scheme for MPSPEs, which is Passive and Partially
Active (PPA). In a PPA scheme, a number of standby nodes will be used
to prepare for recoveries from both independent and correlated
failures. Checkpoints of the processing nodes will be stored at the
standby nodes periodically. Rather than keeping them mostly idled as
in a purely passive approach, we opportunistically employ them for
active replications for a selected subset of the running tasks. In
this way, we can provide very fast recovery for the tasks with active
replicas. Furthermore, when the failed tasks contain those without
active replicas, PPA provides \emph{tentative} outputs with quality
as high as possible. The results can then be rectified after the
passive recovery process has been finished using similar techniques proposed
in~\cite{Balazinska:2008}. In general, PPA is more flexible in utilizing the
available resources than a purely active approach, and in the meantime can
provide tentative outputs with a higher quality than a purely passive
one.  

In this paper, we focus on optimizing utilizing available resources for active
replication in PPA, i.e. deciding which tasks should be included for active
replication. In summary, we have made the following contributions in this paper: 

\textit{\textbf{(1)}} We present PPA, a passive but partially active fault-tolerance
	scheme for a MPSPE. 

  \textit{\textbf{(2)}} As existing MPSPEs often involve user defined
  functions whose semantics are not easily available to the system, we 
  propose a simple yet effective metric, referred to as output fidelity, 
  to estimate the quality of the tentative outputs.
  

  \textit{\textbf{(3)}} We propose an optimal dynamic programming algorithms and
  several heuristic algorithms to determine which tasks to actively replicate
  for a given query topology. 

  \textit{\textbf{(4)}} We implement our approach in an open-source MPSPE, namely
  Storm~\cite{Storm:2014} and perform an extensive experimental
  study on an Amazon EC2 cluster using both real and synthetic datasets. The
  results suggest that by adopting PPA, the accuracy of tentative outputs are
  significantly improved with limited amount of replication resources.

\section{System Model} 
\begin{figure}[h!]
 \centering  
     \includegraphics[scale=0.5]{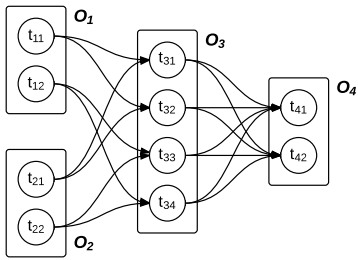}
     \caption{A topology that consists of $4$ operators ($O_1, O_2, O_3, O_4$) with different numbers of tasks.}
    \label{fig:TopologyExample}
\end{figure}

\subsection{Data and Query Model}\label{sec:query_model} 
As in existing MPSPEs~\cite{Storm:2014}, we assume that a data item
is modeled as a key-value pair.  Without loss of generality, the key of a
data item is assumed to be a string and the value is a blob in an
arbitrary form that is opaque to the system.

A query execution plan in MPSPEs typically consists of multiple
operators, each being parallelized onto multiple processing nodes
based on the key of input data.  Each operator is
assumed to be a user-defined function.  
We model such query plan as a topology of the parallel tasks of all
the query operators. By modeling each task as a vertex and the
data flow between each pair of tasks as a directed edge, the
query topology can be represented as a Directed Acyclic Graph (DAG).
Figure~\ref{fig:TopologyExample} shows an example query topology. Each
task represents the workload of an operator that is assigned to a
processing node in the cluster and all the tasks that belong to the
same operator will conduct the same computation.

An operator can subscribe to the  outputs from multiple operators except 
for itself. The output stream of every task will be partitioned into a 
set of substreams using a particular partitioning function,
which divides the keys of a stream into multiple key partitions and splits the
stream into substreams based on these key partitions.  For each task, the input
substreams received from the tasks belonging to the same upstream neighboring
operator will constitute an input stream. Therefore, the number of input streams
of a task is up to the number of its upstream neighboring operators.

Similar to \cite{Zhou:2012}, we consider the following four common
partitioning situations between two neighboring operators in a MPSPE.
In the following descriptions, we consider an upstream operator
containing $N_1$ tasks and a downstream operator containing $N_2$ tasks.
\begin{compactitem}
  \item \emph{One-to-one}: each upstream task only sends data to a single
	downstream task and a downstream task only receives data from a
	single upstream task.
  \item \emph{Split}: each upstream task sends data to $M_2$, $2\leq M_2< N_2$,
	downstream tasks and each downstream task only receives data from
	a single upstream task.
  \item \emph{Merge}: each upstream task sends data to only one downstream
	task and each downstream task receives data from $M_1$, $2\leq M_1< N_1$,
	upstream tasks. 
  \item \emph{Full}: each upstream task sends data to
	all $N_2$ downstream tasks. 

\end{compactitem} 

\subsection{PPA Replication Plan}
Given a topology $T$ and its whole set of tasks $\mathcal{M}$, a PPA replication plan 
for $T$ consists of two parts: a passive replication plan that covers all
the tasks in $\mathcal{M}$ and a partially active replication plan which covers
a subset of $\mathcal{M}$, denoted as $\mathbf{P}$. 
With the passive replication plan, checkpoints will be periodically created 
for all the tasks and stored at the standby nodes. For a task $t_i$, 
its checkpoint consists of $t_i$'s computation state and output buffer. 
After a checkpoint is extracted from $t_i$, its upstream neighboring tasks will be 
notified to prune the unnecessary data from their output buffers. The buffer trimming 
should guarantee that, if $t_i$ fails, its computation state can be recovered by loading 
its latest checkpoint and replaying the output buffers in its upstream tasks. 
On the other hand, for each $t_i \in \mathbf{P}$, an active replica will be created, 
which will receive the same input data and perform the same processing as 
$t_i$'s primary copy.

Upon failures, the actively replicated tasks will be recovered immediately using their 
active replicas, meanwhile the tasks that are only passively replicated
will be restored from their latest checkpoints. When there are some failed tasks
belonging to $\mathcal{M}-\mathbf{P}$, tentative outputs will
be produced before they are fully recovered.
Such tentative outputs have a degraded quality due to 
the loss of input data that otherwise should be processed by the
failed tasks belonging to $\mathcal{M}-\mathbf{P}$. We present how to optimize
the partially active replication plan to maximize the quality of tentative
outputs and the details of the system implementation in the following sections.

\section{Problem Formulation}

\subsection{Quality of Tentative Outputs}
Previous works on load shedding~\cite{Babcock:2004,kang2003evaluating} have
studied how to evaluate the quality of query outputs in case of lost of input data.
Their models assume full knowledge of the semantics of individual operators 
and hence can estimate the output quality in a relatively precise way.
However, in existing MPSPEs, such as Storm, operators are often opaque 
to the system and may contain complex user-defined functions written in
imperative programming languages. The existing models therefore cannot be
easily applied. In our first attempt, we have tried to derive output accuracy
models composed by some generic functions, which should be chosen or
provided by the users according to the semantics of the operators. We
found that this approach is not very user friendly and it may be very difficult for
a user to provide such functions for a complicated operator.

Therefore, we strive to design a model that requires users to provide minimum
information of an operator's semantic, but yet is effective in estimating
the quality of tentative outputs. More specifically, we propose a metric, called
\textit{Output Fidelity $\left( OF\right) $}, which is roughly equal to
the ratio of the source input that can contribute to tentative
outputs. This is based on the assumption that the accuracy of tentative outputs
increases with more complete input and a PPA plan with a higher OF 
value would incur more accurate tentative outputs.

\subsubsection{Operator Output Loss Model}\label{sec:oil} 

\begin{figure}[h!]
 \centering
     \includegraphics[scale=0.4]{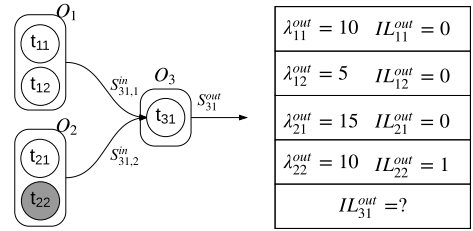}

     \caption{An illustrating topology with task failure,
     where $\lambda_{31,1}^{in} = \lambda_{11}^{out} + \lambda_{12}^{out}$
     and $\lambda_{31,2}^{in} = \lambda_{21}^{out} + \lambda_{22}^{out}$.}
    \label{fig:task_loss}
\end{figure}

It is the sink operator
that produces the final outputs of a topology. As task failures 
can happen at any position within the topology, we need to propagate 
the information losses incurred by any failed task to the output of the sink
operator. Suppose task $t_{22}$ in Figure~\ref{fig:task_loss} is failed,
we need to transform the input loss of $t_{31}$ into its output loss. 
In this subsection, we propose the operator output loss model,  
which estimates the information loss of an operator's output based on the information 
loss of its input. In the next subsection, we present the precise definition of 
OF.

In following descriptions, the set of input streams of task $t_i$ are
denoted as $\left\lbrace S_{i,1}^{in}, S_{i,2}^{in}, ..., S_{i,p}^{in}
\right\rbrace $, where the rate of $S_{i,j}^{in}$ is represented as
$\lambda_{i,j}^{in}$ and its information loss is referred to as $IL_{i,j}^{in}$.
The rate of $t_i$'s output stream, $S_i^{out}$, is referred to as
$\lambda_{i}^{out}$, and its information loss is denoted as $IL_i^{out}$. If $t_i$
is failed, its output will be lost and $IL_i^{out}$ will be set as $1$.
Otherwise, we calculate $IL_i^{out}$ based on the information losses of
$t_i$'s input streams. 

As described in the query model, an input stream of a task may consist of multiple
substreams, which are sourced from tasks belonging to the same upstream
neighboring operator. Suppose that $S_{i,j}^{in}$ consists of a set of substreams 
$U_{i,j}^{in}$. For each substream $s_k$, $s_k \in U_{i,j}^{in}$, denoting
its rate as $\lambda_{s_k}$ and its information loss as $IL_{s_k}$,
then the information loss of $S_{i,j}^{in}$ is calculated as: 
 \vspace{-5pt}
\begin{equation}
IL_{i,j}^{in} = \frac{\sum_{s_k}^{s_k \in U_{i,j}^{in}} \lambda_{s_k} \cdot IL_{s_k}}
{\sum_{s_k}^{s_k \in U_{i,j}^{in}} \lambda_{s_k}}
\end{equation}

Meanwhile, the output stream of task $t_i$, $S_{i}^{out}$, can be split into a
set of substreams, denoted as $D_{i}^{out}$. For each substream $s_k$ belonging
to $D_i^{out}$, its information loss is estimated to be equal to $S_{i}^{out}$,
i.e. $IL_{s_k} = IL_{i}^{out}$.

Figure~\ref{fig:task_loss} depicts an example topology as well as the 
rate of each output stream. $IL_{31}^{out}$ represents the information 
loss of output stream $S_{31}^{out}$ caused by the failure of task $t_{22}$. 
We distinguish two situations and use this example to illustrate the 
calculation of information loss of a task's output stream.

{\bf Correlated-Input Operator}. $O_t$ performs computations over the join
results of its input streams. For example, suppose $O_3$ in
Figure~\ref{fig:task_loss} is a join operator. Without further
semantic information of $O_3$, we consider the effective input of $t_{31}$ as the
Cartesian product of its input streams, whose rate is equal to
$\left( \lambda_{31,1}^{in} \cdot \lambda_{31,2}^{in}\right) $ 
and its information loss can be computed as
 $\left[ 1 - \left( 1- IL_{31,1}^{in}\right)  \cdot \left( 1 - IL_{31,2}^{in}\right) \right] $.  
By assuming that the information loss of $t_{31}$'s output 
should be equal to that of its input stream, we can get
$IL_{31}^{out} = \frac{2}{5}$. 
In summary, the information loss of $t_i$'s output stream 
can be calculated as:

 \begin{equation} 
 IL_{i}^{out}=1 - \prod_{j=1}^{p} \left( 1 - IL_{i,j}^{in}\right)
 \end{equation} 

{\bf Independent-Input Operator}. $O_t$ does not compute joins over input
streams. If $O_3$ in Figure~\ref{fig:task_loss} is an independent-input
operator, the effective input of $t_{31}$ is considered as the union of
its input streams, whose rate is equal to 
$\left( \lambda_{31,1}^{in} + \lambda_{31,2}^{in}\right) $ and its input loss
can be calculated as $\frac{\lambda_{31,1}^{in} \cdot IL_{31,1}^{in} + 
\lambda_{31,2}^{in} \cdot IL_{31,2}^{in}}{\lambda_{31,1}^{in} + \lambda_{31,2}^{in}}$.
Similar to the correlated-input operator, we also assume that
the information loss  of $t_{31}$'s ouptut 
should be equal to that of its input stream.
Then we have, in this example,  $IL_{31}^{out} = \frac{1}{4}$. 
In general, the information loss of $t_i$'s output stream can be calculated 
as follows: 
 \begin{equation} 
IL_{i}^{out} =
\frac{\sum_{j=1}^{p} \lambda_{i,p}^{in} \cdot IL_{i,p}^{in}}{\sum_{k=1}^{p} \lambda_{i,k}^{in}}  
 \end{equation} 
 
Recall that one of the design principles is to request as little information of
the operators' semantics as possible. We distinguish the aforementioned two
types of operators simply because the characteristics of their effective inputs are
very different. With such distinction, the OF metric can be estimated much more
precisely. 


\subsubsection{Output Fidelity}

With the operator output loss model,
the output information losses of tasks in the sink operator
can be calculated by conducting a depth-first traversal of the topology, which
starts from the tasks in the source operators and ends at the tasks in the sink
operator.

By denoting the sink operator of topology $T$ as $O_{sink}$, 
and the set of tasks belonging to $O_{sink}$ as 
$\left\lbrace t_1, t_2, ..., t_{M_t} \right\rbrace$, 
The output fidelity of topology $T$, $OF_{T}$, is defined as: 
 \begin{equation}
OF_{T} =  1- \frac{\sum_{i=1}^{M_t} \lambda_{i}^{out} \cdot IL_{i}^{out}}{\sum_{j=1}^{M_t} \lambda_{j}^{out}}
\label{eqn:ilr}
 \end{equation}

\vspace{-10pt}
\subsection{Problem Statement}\label{problem} 

Before presenting the problem definition, 
we introduce a concept: \textit{Minimal Complete Tree}, 
which is also referred to as \textit{MC-tree} for simplicity in the 
following sections.

\newtheorem{mydef}{Definition}
\begin{mydef}\label{def1}
\textsc{Minimal Complete Tree (MC-Tree)}: A minimal complete tree is a 
tree-structured subgraph of the topology DAG. The source vertices of this 
subgraph correspond to tasks from the source operators and its sink vertex 
is a task from an output operator. A minimal complete tree can continuously
contribute to final outputs if and only if all its tasks are alive.
\end{mydef}

Taking the topology in Figure~\ref{fig:TopologyExample} for instance, 
if $O_3$ is an independent-input operator, tasks in
$\left\lbrace t_{11}, t_{31}, t_{41} \right\rbrace $ can constitute an MC-tree
and there are in total $16$ MC-trees in the topology.
However, if $O_3$ is a correlated-input operator, $t_{31}$ cannot produce
any output if either $t_{11}$ or $t_{21}$ fails. Hence tasks in
$\left\lbrace t_{11}, t_{21}, t_{31}, t_{41} \right\rbrace $ can constitute
an MC-tree and the number of MC-trees in the topology is equal to $8$. 

Based on Definition~\ref{def1}, if failures of tasks in an MC-Tree
occur, it will only continue propagating data to the sink operator if and
only if all of its failed tasks are actively replicated. 
Suppose topology $T$ consists of a set of operators $O_1, O_2, ..., O_N$ 
and the available resources can be used to actively replicate $R$ tasks
($R \leq |\mathcal{M}|$, where $\mathcal{M}$ is all the tasks of $T$), then the
problem of optimizing a partially active replication plan is defined as follows:  

\begin{mydef}\label{def2}
\textsc{Partially Active Plan}: Given a query topology $T$,
choose $R$ tasks for active replication such that, the output fidelity of
the partial topology that is composed of the actively replicated 
MC-trees in $T$ is maximized.
\end{mydef}
This problem is NP-hard, as it can be polinomially reduced from the Set-Union Knapsack Problem~\cite{NAV:NAV3220410611}, 
which is NP-hard.

\section{Active Replication Optimization}
Recall that we consider the worst case scenario for a correlated
failure, i.e. there is at least one failed task in every MC-tree.  
Before the completion of the
passive recovery process, only the MC-trees whose failed
tasks are actively replicated can produce tentative outputs. 
The optimization objective is to maximize the value of OF
with limited amount of resources used for active
replication. 

\vspace{-5pt}
\subsection{Dynamic Programming}\label{sec:dyna}
We first present a dynamic programming algorithm that can generate an
optimal replication plan for correlated failure. As has been introduced
in section~\ref{problem}, we take MC-tree as the basic unit for 
replication candidates in the algorithm.
\begin{algorithm}[t]
\small
\caption{Dynamic Programming: \textsc{PlanCorrelatedFailure}($R$)}
\label{alg:danymic}
\KwIn{ Amount of available resources $R$\;}
\KwOut{Replication plan $\mathbf{P}$\;}
$CP_0\leftarrow \emptyset$; $usage \leftarrow 0$; $SC\leftarrow \{CP_0\}$\; 
\tcc{$CP_0$:initial replication plan; $SC$:candidate plan set;} 

\While{$ usage++ < R $}
{
	\ForEach{candidate plan $CP_i$ in $SC$}
	{
		$dif \leftarrow usage - |CP_i|$\; \tcc{$|CP_i|$ is the number
		  of replicated tasks in $CP_i$ and $dif $ is the number of tasks that can be
		added to $CP_i$ at this step;}
		$UT_i \leftarrow \{$ MC-tree $tr \mid tr \notin CP_i\}$\;
		$u_{i} \leftarrow $ $\max\{\mathbf{nonrep\_tasks}(tr, CP_i) \mid tr \in UT_i\}$\;
		\tcc{$\mathbf{nonrep\_tasks}(tr, CP_i)$ returns the number of
	  non-replicated tasks of MC-tree $tr$ in $CP_i$;}
		\If{$dif \leq u_{i}$}
		{
		  \ForEach{$\textrm{MC-tree } tr_j\in \{tr\mid tr\notin CP_i ~ \& ~ \mathbf{nonrep\_tasks}(tr_j, CP_i) == dif\}$}
			{
				$CP_j \leftarrow CP_i \cup tr_j$\;
				\If{$CP_j \notin SC$}{
				 Add $CP_j$ to $SC$\;
				}
			}
		}
		\lElse{Remove $CP_i$ from $SC$\;}
	}
}
$\mathbf{P} \leftarrow $ the candidate plan in $SC$ with the maximal OF value.
Return $\mathbf{P}$;
\end{algorithm}
Details of this algorithm are presented in
Algorithm~\ref{alg:danymic}. It is essentially a bottom-up dynamic
programming algorithm.  We incrementally increase the number of
resources to be used for active replication and enumerate the possible
expansions of the plans produced in the previous step. Assuming the minimum
size of MC-trees is $r$, one can obtain the first set of replication
plans, referred to as $SC$, by replicating $r$ tasks.  At this step,
each plan in $SC$ contains exactly one MC-tree.  Note that the MC-trees that
have not been added to a candidate plan $CP_i$ may also have
replicated tasks if they share some tasks with another MC-tree within
$CP_i$.

At the next iteration of the while loop starting at line $2$, we
increase the resource usage by $1$.  We scan through each candidate
plan $CP_i \in SC$ to see if there is an MC-tree $tr_j\notin CP_i$ that
contains a number of non-replicated tasks which is equal to $usage - |CP_i|$,
where $|CP_i|$ is the number of replicated tasks in $CP_i$. For each MC-tree
satisfying this condition, we create a new candidate plan $CP_j$
(line $9$) such that $CP_j \leftarrow CP_i \cup tr_j$.  If $CP_j$ has no
duplicate in $SC$, then it will be inserted into $SC$. 
The algorithm will continue until $usage$ is equal to the limit $R$.

The cost of scanning through $SC$ can be reduced by removing a
candidate plan $CP_i$ from $SC$ if all its possible expansions have been
considered. More precisely, remove $CP_i$ from $SC$ if the
maximum number of non-replicated tasks of the MC-trees not included in
$CP_i$ is less than the difference between the available resource at
the current iteration, i.e. $usage$, and the current number of
replicated tasks in $CP_i$ (lines $7$ and $12$). 
After the while loop is finished, the candidate plan with the maximal OF
in $SC$ will be returned.

The upper bound of the complexity of this algorithm is $O\left( 2^\mathcal{T}\right)$,
where $\mathcal{T}$ is the number of MC-trees in the query topology, which varies 
with the topology structures and has an upper bound of $O(M^N)$, 
where $N$ is the number of operators and $M$ is the average degree of parallelization of
operators in $T$.
The following theorem states the optimality of this dynamic 
programing algorithm, the proof is skipped due to space limitation.
\newtheorem{thm2}{Theorem}\label{th}
\begin{thm2}
  Let $\mathbf{P}$ be the replication plan produced
  by Algorithm~\ref{alg:danymic} and $\mathbf{P_t}$ be a different
  replication plan. If $OF_{P_t} \geq OF_{P}$, then the resource usage of 
  $\mathbf{P}$ is always equal to or less
  than that of $\mathbf{P_t}$.  
\end{thm2}

\begin{algorithm}[t]

  \small
  \caption{\textsc{Greedy}($R$)}
\label{alg:independent}
\KwIn{
The amount of available resources $R$\;
}
\KwOut{Replication plan $\mathbf{P}$}
Initialize: $AS \leftarrow \emptyset$\;
\ForEach{Task $t_i \notin \mathbf{P}$}{
				$A_{i}\leftarrow $ the value of OF if only $t_i$ fails\;
				$AS \leftarrow AS \cup \{A_{i}\}$\; 
				}
Sort $AS$ in ascending order\;
$TS \leftarrow$ set of tasks whose corresponding OF values are 
among top-$R$ in $AS$\;
$\mathbf{P} \leftarrow \mathbf{P} \cup TS$\;
Return $\mathbf{P}$
\end{algorithm}

\subsection{Greedy Algorithm}

We present a greedy algorithm. For each task in the topology,
the greedy algorithm will calculate the OF of the topology
by only failing this task. A task whose failure would lead to a smaller
OF  will be assigned a higher priority for replication.
We present the details of this greedy algorithm in Algorithm
\ref{alg:independent}, which will first rank all the tasks in ascending order
based on the OF  calculated by their respective failures. Then
it will iterate to choose the corresponding task that would cause the
minimal OF  among all the remaining non-replicated tasks
in the set $AS$. 

The complexity of the greedy algorithm is 
$O(N \cdot M)$, where the notations are defined in Section~\ref{sec:dyna}.
Although this complexity is much lower than 
that of the dynamic programming algorithm, it fails to consider 
whether the tasks in the replication plan could form complete MC-trees, which
will damage its performance especially when the number of active replicated tasks
is small. The experimental results in section~\ref{sec:tentative} can verify
this defect of the greedy algorithm.

\subsection{Structure-Aware Algorithm}
The dynamic programming
algorithm searches for the optimal plan by selecting a subset of MC-trees
for replication under the resource constraint to maximize the value of OF.
Inspired by this, we design a structure-aware algorithm
that, at each step, rather than enumerating all the possible
expansions of a candidate plan, only expands it with an MC-tree that can
incur the greatest increase in OF per resource unit. 

Unfortunately, even such a greedy approach  may fall short under the
following situation. Consider a topology $T$ that consists of a
sequence of $k$ operators and all the operators use Full partitioning, 
the number of MC-trees within $T$ is equal
to $\prod_{i=1}^k M_i$, where $M_i$ is the number of tasks of operator
$O_i$. In such a topology, the number of MC-trees 
will grow very fast with increasing number of operators. 
Therefore, even a greedy search among the
possible combinations of MC-trees would not perform well. 

To solve this problem,  we firstly decompose a general topology
into two  types of topologies, namely full topologies and structured
topologies, and then optimize them separately. The definitions of these two types of
topologies are as follows: 

\begin{compactitem}
\item \textbf{Structured topology} is defined as a topology where
only the operators, that produce outputs of this
topology, can have a Full partitioning function and the others have
other types of partitioning functions. 
\item \textbf{Full topology} is defined as a topology that
  all of its operators have a Full partitioning function.
  \end{compactitem}

The rest of this section is organized as follows: firstly, we present
the algorithms generating PPA plans for structured topologies and
full  topologies respectively. Then we will explain the structure-aware algorithm,
which generates the PPA plan for a general topology by decomposing it into several
sub-topologies, each being either a structured topology or a full topology.

%

\subsubsection{Algorithm for Structured Topology} \label{sec:structure}
\begin{algorithm}[t]
  \small
  \caption{\textsc{PlanStructuredTopology}($\mathbf{P}, R, T$) }
\label{alg_structured}
\KwIn{
  An initial plan $\mathbf{P}$;
  The amount of available resources $R$;
  Topology $T$;
}

\KwOut{Replication plan $\mathbf{P}$\;}

$usage = 0$;
$S_u \leftarrow$ Set of the units split from topology $T$\;

\ForEach{Unit $U_i \in S_u$}{
	Build segment set $G_i$\;
}

\While{$usage \le R$}{
  $Candidates\leftarrow \emptyset$ \;
\ForEach{Unit $U_i \in S_u$}{
	\ForEach{non-replicated segment $g_i \in U_i$}{
	  $CG_{i} \leftarrow \{g_i\}$\;
	  \If{${OF_{\mathbf{P}}} = {OF_{\mathbf{P} \cup CG_i}}$ }
		{
			Conduct a BFS from $U_i$ to traverse all the units:\\
			\ForEach{visited unit $U_j$ during the BFS}{
			Segment $g_j \leftarrow \mathbf{max\_of}\left( U_j \right)$ \;
			\tcc{$\mathbf{max\_of}\left( U_j \right)$ returns the segment in $U_j$,
			 which is connected with segment in $CG_i$ and has the maximal OF with $U_j$ treated as 
			 an independent topology;}
			 \If{$|CG_i| + |g_j| \leq usage$}
			 {$CG_i = CG_{i} \cup g_{j}$\;}
			 \lElse{Stop the BFS\; }
			}
		}
		$Candidates \leftarrow Candidates \cup CG_{i}$\; 
	}
}

Find $CG_{opt}$ from $Candidates$ such that the following value is maximized:
$({OF_{\mathbf{P} \cup CG_{opt}} -
OF_{\mathbf{P}}})/{|CG_{opt}|}$\; 
$\mathbf{P}  = \mathbf{P} \cap CG_{opt}$;
$usage = usage + |CG_{opt}|$\;

\lIf{$CG_{opt} \neq \emptyset$}{return $\mathbf{P}$\;}

Remove the completely replicated units from $S_u$;
}
Return $\mathbf{P}$\;		
\end{algorithm}

Although we define structured topology such that Full partitioning
only exists in the output operators, the number of MC-trees in a
structured topology could still be very large. Consider the situation
that a task $t_i$ receives $N_{in}$ input streams and produce
$N_{out}$ output streams, there will be at least $N_{in} * N_{out}$ MC-trees
containing $t_i$. In addition,  if $t_i$ joins
$N_k$ substreams from operator $O_k$ with $N_j$ substreams 
from operator $O_j$, the number of MC-trees containing $t_i$
will at least be equal to  $N_k \cdot N_j$.
To avoid bad performance due to the large
number of MC-trees, we split a structured topology into multiple units 
such that, within a unit, the number of MC-trees is equal to the
maximal number of input substreams among the operators of this
unit. We refer to an MC-tree in a unit as \emph{segment} to differentiate
it from the concept of a complete MC-tree in the topology. 

\begin{figure}[htb]

    \centering
    \subfigure[]{
        \label{fig:seg_1}
        \centering
        \includegraphics[width=0.45\linewidth]{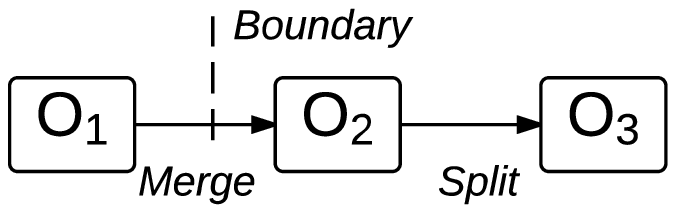}
    }
    \hspace{-5pt}
    \subfigure[]{
        \label{fig:seg_2}
        \centering
        \includegraphics[width=0.45\linewidth]{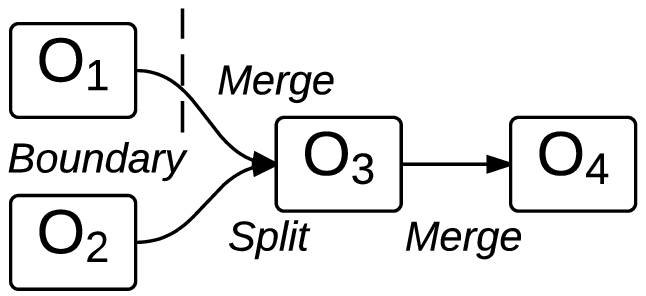}
    }
    \caption{Examples of splitting structured topologies into units. $O_3$ in 
    Figure~\ref{fig:seg_2} is a join operator.}\label{fig:segment}

\end{figure}

The situation of multiple input streams and multiple output
streams occurs on the task who has an input stream partitioned with 
Merge and an output stream partitioned with Split, 
a unit boundary will be set between this operator and its upstream neighboring
operator using Merge partitioning. For instance, a unit boundary is set between
$O_1$ and $O_2$ in the topology in Figure~\ref{fig:seg_1}.
The situation that a task joins multiple input substreams from one
operator with substreams from other operators happens on the tasks of join operators
that have at least one input stream partitioned with Merge.
As illustrated in Figure~\ref{fig:seg_2}, a unit boundary is set between 
$O_1$ and $O_3$.

Note that, with such a decomposed topology, replicating a segment is
beneficial only if all the other segments within the same complete
MC-tree are also replicated. In other words, we should avoid enumerating
plans that replicate a set of disconnected segments.

The details of the algorithm for structured topology are
presented in Algorithm \ref{alg_structured}. The algorithm searches
through the units generated from input topology.  Within unit
$U_i$, if the set of non-replicated segments is not empty, we check
whether replicating these segments will increase the final output
accuracy (line $9$). Note that this will only be true if this segment
can form a complete MC-tree with the other replicated segments within the
current plan. Each of such segments will be put into a candidate pool
(line $16$). If the segment $g_i$ does not enhance the plan's OF,  we
conduct a BFS (Breadth-first search) starting from $U_i$ and
traversing through all the units in Topology T.  The BFS is terminated
until $usage$ is less then the non-replicated tasks in $CG_i$.
Finally, every unit visited during the BFS contributes a segment to
$CG_i$ and the segments from neighboring units are connected (lines $10 - 15$).
Then we put such a set of segments as one candidate in the
candidate pool. 

After finishing the scanning of all units, we get a candidate pool
consisting of a number of segment sets, each containing one or more
segments. We use a profit density function to rank the candidates. The
profit density of a candidate $CG_k$ is calculated as
$\left. \left( OF_{P \cup CG_k} - OF_{P} \right) \middle / |CG_k|\right.$,
where $OF_{P}$ is the OF value of plan $P$, $OF_{P \cup
CG_k}$ is the OF value after expanding $P$ by replicating segment in
$CG_k$. $|CG_K|$ is the number of non-replicated tasks within
$CG_k$. The plan in the candidate pool with the maximum profit
density will be merged with the input plan $P$ and returned.
The complexity of Algorithm \ref{alg_structured} is equal to
$O(R \cdot N \cdot M^2 \cdot E)$, where $R$ is the amount of available replication
resources, $N$ is the number of operators, $M$ represents the
average degree of parallelization of operators in $T$, and $E$ is the number of 
neighboring unit pairs.
   
 \begin{algorithm}[t]
  \small
  \caption{\textsc{PlanFullTopology}($\mathbf{P}$, $R$, $T$)}
\label{alg:optimal_full}
\KwIn{
  Initial replication plan $\mathbf{P}$;
  Amount of available resources $R$;
  Topology $T$;
  
}
\KwOut{Replication Plan $\mathbf{P}$}
Initialize:
		$usage \leftarrow 0$\;
		$N \leftarrow \; Number \; of \;operators$\;
		Sort the set of tasks $S_i$ of each operator $O_i$ based on the
		OF increase, $\delta_{ij}$ , of tasks\;
\If{$P = \emptyset \; \& \; N \leq R$}{
	\ForEach{$O_i$}{
	Let $p_{ik}$ be the node in $S_i$ that has the largest OF increase $\delta_{ik}$\;
		$\mathbf{P}\leftarrow \mathbf{P} \cup \{p_{ik}\}$; 
		$S_i \leftarrow S_i - \{p_{ik}\}$\;
	}
	$usage = N $\;
}
\lIf{$\mathbf{P} = \emptyset \; \& \; N > R$}{return $\mathbf{P}$\;}
\While{$ usage < R$}{
  	$Candidates \leftarrow \emptyset$\;
	\ForEach{$O_i$}{
		Let $p_{ik}$ be the node in $S_i$ that has the largest OF increase $\delta_{ik}$\;
		$Candidates \leftarrow Candidates \cup P_i \cup \{p_{ik}\} $\;
	}
	$P_j \leftarrow \mathrm{\mathbf{max\_accuracy\_plan}}(Candidates)$\;
	$S_j \leftarrow S_j - \{p_{jk}\}$;
	$\mathbf{P} \leftarrow P_j$; $usage++$\;
}
Return $\mathbf{P}$\;
\end{algorithm}

\subsubsection{Algorithm for Full Topology}

Each task within a full topology will send input data to
all the tasks that belong to its downstream neighboring operators.  
We propose an algorithm for full topology as illustrated
in Algorithm~\ref{alg:optimal_full}. The basic idea of this
algorithm is that, within any operator, we always prefer to replicate
the task that will bring the maximum increase of OF under the 
assumption that all the other tasks that belong to the same operator
are failed and the tasks that belong to other operators are alive.
We denote the increase of OF by replicating task $t_{ij}$ as
$\delta_{ij}$.  If the input plan $P$ is empty, we first select one
task from each operator that has the largest $\delta_{ij}$ among all the
tasks in this operator and put it into $P$ (lines $4-7$). If $P$ is not
empty, we iterate and select $R$ tasks that have larger OF increases, 
i.e. $\delta_{ik}$ , than other tasks in the topology 
and put them into $P$ (lines $10-16$).  
The complexity of this algorithm is $O(N \cdot R)$, where
$R$ is the amount of available replication
resources and $N$ is the number of operators.

{\renewcommand\baselinestretch{0.8}
\begin{algorithm}[t]
\small
  \SetKwFunction{PlanSubTopology}{PlanSubTopology}
  \SetKwInput{Func}{Function} 
\caption{\textsc{StructureAware}($R$,$T$)}
\label{alg:general_correlated}
\KwIn{The amount of available resources $R$; Topology $T$\;}
\KwOut{Partial replication plan $\mathbf{P}$\;}

	Initialize: decompose the complete topology $T$ into
	sub-topologies: ${TS_1, TS_2, ...}$ \;
	$\mathbf{P} \leftarrow \emptyset$, $S_A \leftarrow \emptyset$, $usage \leftarrow  0$\;
	\If{$R <$  Number of operators in $T$}{
		Return $\mathbf{P}$ \;	
	}
	
	\ForEach{Sub-Topology $TS_i$}{
			$N_i \leftarrow$ Number of operators in $TS_i$\;
			$P_i \leftarrow$ \PlanSubTopology($\emptyset, R_i, TS_i$); 
			$\mathbf{P} \leftarrow \mathbf{P} + P_i$\;
			$P_i^{\prime} \leftarrow$ \PlanSubTopology($P_i, R_i, TS_i$)\;
			
			$C_i \leftarrow |P_i^{\prime}| - |P_i|$;
			$\Delta_i \leftarrow \frac{OF_{P_i^{\prime}} - OF_{P_i}}{C_i}$\;		
			Put $\Delta_i$ into $S_A$ in descending order;
			$usage += N_i$\;
		}
	\While{$usage < R$}{
		$LastUsage \leftarrow usage$;
		$j \leftarrow 1$\;
		\While{$j \le |S_A|$}{
			$\Delta_i \leftarrow$ $j$th value in $S_A$; $j++$\;	
			\If{$C_i + usage \leq R$ }{
				Use $P_i^{\prime}$ to replace $P_i$ in $\mathbf{P}$\;
				Calculate new $C_i$, $\Delta_i$. Insert $\Delta_i$ into $S_A$ in descending order; 
				break\;
			}
				 
		}
		\lIf{$usage = lastUsage$}{break\;}
	}
	Return $\mathbf{P}$\;
  
  	\Indm \Func{\PlanSubTopology{$\mathbf{P}$, $N_i$, $T$}}
    \Indp

	\If{$T$ is a full topology}
	    { $\mathbf{P} \leftarrow$ \textsc{PlanFullTopology}$(\mathbf{P}, N_i, T) $\; }
		\lElse{ $\mathbf{P} \leftarrow$ \textsc{PlanStructuredTopology}$(\mathbf{P}, N_i, T) $\; }

\end{algorithm}
}

\subsubsection{Solution for General Topology}

With the above algorithms for specific topology structures, we divide
a general topology into several sub-topologies and then use the
corresponding algorithms according to the type of each sub-topology to
generate the replication plans. We require that at least one partitioning 
function between any two neighboring sub-topologies is Full and the amount of
sub-topologies is minimized. The reason behind this requirement is to
make the selection of the replication segments in the sub-topologies 
independent from each other. 

\begin{figure}[htb]
 \centering
     \includegraphics[scale=0.6]{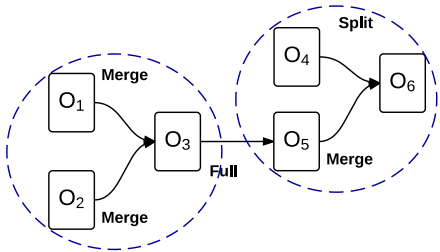}
     \caption{Example of splitting a topology into sub topologies.}
    \label{fig:subtopology}
    
\end{figure}

The split algorithm explores the topology using multiple depth-first
searches (DFS). At the beginning, only the sink operator of the
given topology is in the start point set $SP$. At each iteration, we
will pick an operator, $O_s$, from $SP$ and build a sub-topology by
performing a DFS starting from $O_s$. If the DFS arrives at an
operator $O_i$ whose partitioning function is incompatible with the
type of the current sub-topology, it will not further traverse $O_i$'s
downstream operators and $O_i$ will not be added to the current
sub-topology but instead be put into $SP$. Finally the algorithm will
terminate until $SP$ is empty.  Figure~\ref{fig:subtopology} presents
an example general topology, which is decomposed into two
sub-topologies: $\{O_1, O_2, O_3\}$ and $\{O_4, O_5, O_6\}$.

We present details of the correlated-failure optimization algorithm
for a general topology in Algorithm~\ref{alg:general_correlated}, which is
referred to as the Structure Aware algorithm.  The
algorithm first decomposes the topology into sub-topologies
which are either full topologies or structured
topologies. Then the algorithm runs in multiple iterations. Within
each iteration, it will try to get a replication plan from each
sub-topology and select the one with the maximum profit density (lines
$11 - 17$). The loop will be terminated when there is no more resource to
replicate a complete MC-tree. The algorithm's complexity is equal to
$O(R \cdot N \cdot M^2 \cdot E)$,
where the notations are defined in Section~\ref{sec:structure}.


\section{System Implementation}\label{system}

\subsection{Framework}

\begin{figure}[h!]
 \centering
     \includegraphics[scale=0.5]{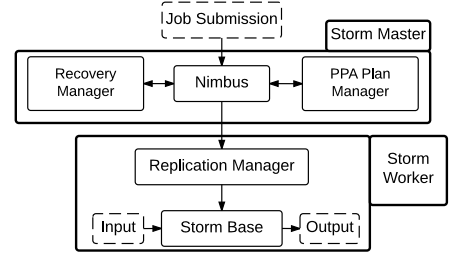}
     \caption{System Framework}
    \label{fig:system}
\end{figure}


We implemented our system on top of Storm. In comparing to Spark Streaming,
which processes data in a micro-batching approach, Storm will process an input
tuple once it arrives and thus can achieve sub-second end-to-end processing
latency. As shown in Figure~\ref{fig:system}, the nimbus in the Storm master
node assigns tasks to the Storm worker nodes and monitoring the failures. On
receiving a job, the nimbus will transfer the query topology to the PPA plan
manager, which will generate a PPA recovery plan under the constraint of
resource usage of active replication. The PPA recovery plan  consists of two
parts: a completely passive standby plan and a partially active replication
plan.  Based on the PPA recovery plan, the replication manager in the worker
nodes will create checkpoints to passively replicate the whole query topology.
Checkpoints will be stored onto a set of standby nodes.  The replication manager
will create active replicas for the tasks that are included in the partially
active replication plan. The active replicas can support fast failure recovery
and will also be deployed onto the standby nodes.

Once a failure is detected by the nimbus, The recovery manager in the Storm master node
will decide how to recover the failed tasks based on the PPA replication plan.
For the tasks that are actively replicated, the recovery manager will notify the
nimbus to recover them using their active replicas such that the tentative results could
be produced as soon as possible. The failed tasks that are passively replicated
 will be  recovered with their latest checkpoints.

\subsection{PPA Fault Tolerance}

\textbf{Passive Replication.} In PPA, checkpoints of the processing tasks will
be periodically created and stored at the standby nodes. 
We adopted the batch processing approach~\cite{Zaharia:2013} to guarantee
the processing ordering of inputs during recovery is identical to that
before the failure. With this approach, input tuples are divided into a
consecutive set of batches. A task will start processing a batch after it
receives all its input tuples belonging the current batch. This is ensured by
waiting a batch-over punctuation from each of its upstream neighboring tasks.
Tuples within a batch will be processed in a predefined round-robin order. The
effect of batch size on the system performance has been researched in previous
work~\cite{Das:EECS-2014-133}.

A single point failure can be recovered by restarting the failed task, loading
its latest checkpoint and replaying its upstream tasks' buffered data. The
downstream tasks will skip the duplicated output from the recovering task until
the end of the recovery phase. While recovering a correlated failure, if a task
and its upstream neighboring task are failed simultaneously and its checkpoint
is made later than its upstream peers', the recovery of the downstream task can
only be started after its upstream peer has caught up with the processing
progress. In other words, synchronizations have to be carried out among the
neighboring tasks.


\textbf{Active Replication.} 
If task $t$ has an active replica $t^{\prime}$, 
the output buffer of $t^{\prime}$ will store the output tuples produced
by processing the same input in the same sequence as $t$ does. 
The downstream tasks of $t$ will subscribe the outputs from both 
$t$ and $t^{\prime}$. By default, the
output of $t^{\prime}$ is turned off. To reduce
the buffer size on $t^{\prime}$, its primary, $t$, will periodically
notify $t^{\prime}$ about the latest output progress and the latter can then
trim its output buffer. If $t$ is failed, $t^{\prime}$ will
start sending data to the downstream tasks of $t$. The downstream tasks will
eliminate the duplicated tuples from $t^{\prime}$ by recognizing their sequence
numbers. The batch processing strategy can guarantee an identical processing 
order between the primary and active replica of a task.

\textbf{Tentative Outputs.}
As checkpoint-based recovery requires replaying the buffered data and
synchronizations among the connected tasks and hence incurs significant recovery
latency, PPA has the option to continue producing tentative results once the
actively replicated tasks are recovered. Recall that during normal processing, a
task will only start processing a batch after receiving the batch-over
punctuations from all of its upstream neighboring tasks. If any of its upstream
neighboring tasks fails, the recovery manager in the Storm master node will
generate the necessary batch-over punctuations for those failed tasks, such that
a batch could be processed without the inputs from the failed tasks and tentative
outputs will be generated with an incomplete batch.  After the failed tasks are
recovered, the recovery manager will stop sending the batch-over messages for
them such that the downstream tasks will wait for the batch contents from the
recovered tasks before processing a batch.  After all the failed tasks are
recovered, the topology will start generating accurate outputs. 

In this paper, we assume the adoption of similar techniques proposed
in~\cite{Balazinska:2008} to reconcile the computation state and correct
the tentative outputs and leave the implementation of these techniques as our
future work. 
\subsection{Dynamic Plan Adaptation}
Considering that tasks' input rates may fluctuate over time, the active replication
plan should be dynamically adapted accordingly. The PPA plan manager periodically 
collects the input rates of all the processing tasks and generate new active replication
plan. If the new plan is different from the previously applied plan, 
applying the new plan may require deactivating the active replicas of a set of tasks 
and generating active replicas for another set of tasks. Deactivating the active replicas 
can be implemented by terminating their processing and releasing their occupied
resources. To generate new active replicas, we can send the corresponding 
checkpoints to the destination nodes and initialize the state of the active
replicas by using the checkpoints. The newly started active replicas will receive the
buffered outputs from their upstream neighboring tasks and then start the
processing.  Eventually, the newly generated active replicas will catch up with
the progress of their primary copies. Dynamic plan adaptation is not implemented
in the current version of our system, which is part of our future work. 

\section{Evaluation}

The experiments are run over the Amazon EC2 platform. We build a cluster
consisting of 36 instances, of which 35 m1.medium
instances are used as the processing nodes and one c1.xlarge
instance is set as the Storm master node.
Heartbeats are used to detect node failures in a 5-second interval. 
The recovery latency is calculated as the time interval between 
the moment that the failure is detected and the instant when the failed task
is recovered to its processing progress before failure. The processing 
progress of a task is defined as a vector.  Each field of the progress vector
contains the sequence number of the latest processed tuple from a specific 
input stream of the task. A failed task is marked as recovered if the 
values of all the fields in its current progress vector are larger than or equal to
the values of the corresponding fields of the progress vector before failure.
Additional information of the experiment configuration will be presented in the
following sections.

\subsection{Recovery Efficiency}

\begin{figure}[h!]
 \centering       \vspace{-5pt}
     \includegraphics[scale=0.65]{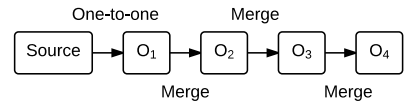}

     \caption{Topology used in the 
     experiments of recovery efficiency in the scale of operator.}
    \label{fig:stopo}
\end{figure}

  \begin{figure*}[htb]
      \centering
      \begin{minipage}{.33\textwidth}
          \includegraphics[scale=0.5]{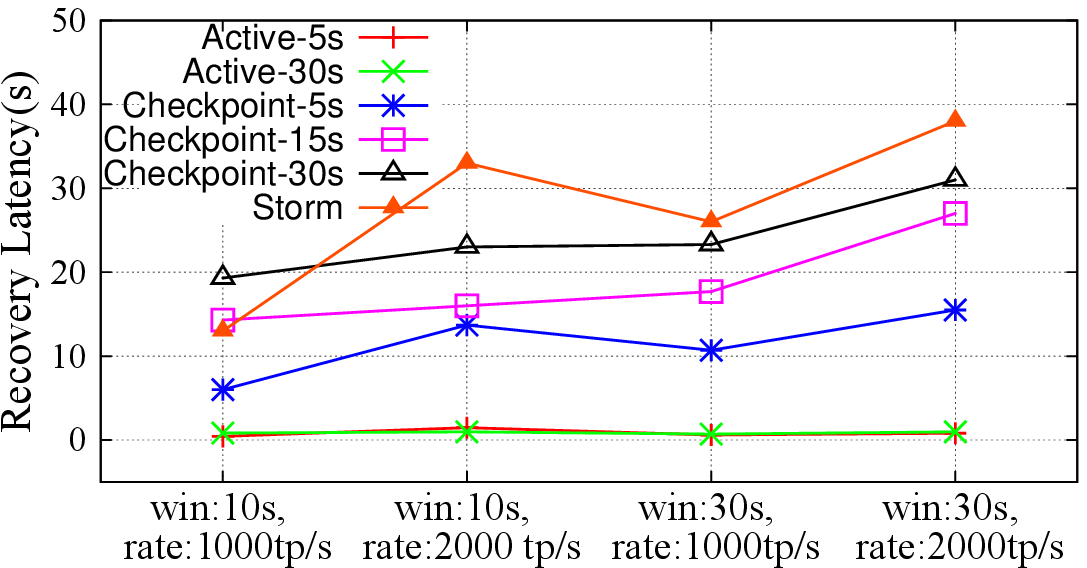}
          \caption{Recovery latency of single node failure.}
          \label{fig:single}
      \end{minipage}%
      \begin{minipage}{0.33\textwidth}
          \includegraphics[scale=0.5]{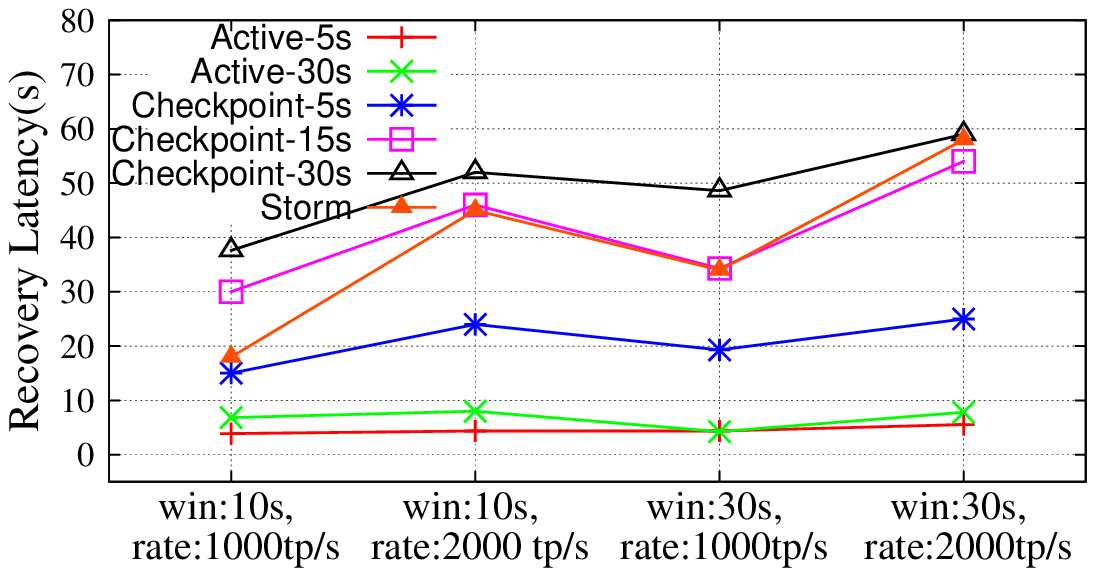}
          \caption{Recovery latency of correlated failure.}
          \label{fig:correlated}
      \end{minipage}%
      \begin{minipage}{.33\textwidth}
                      \includegraphics[scale=0.5]{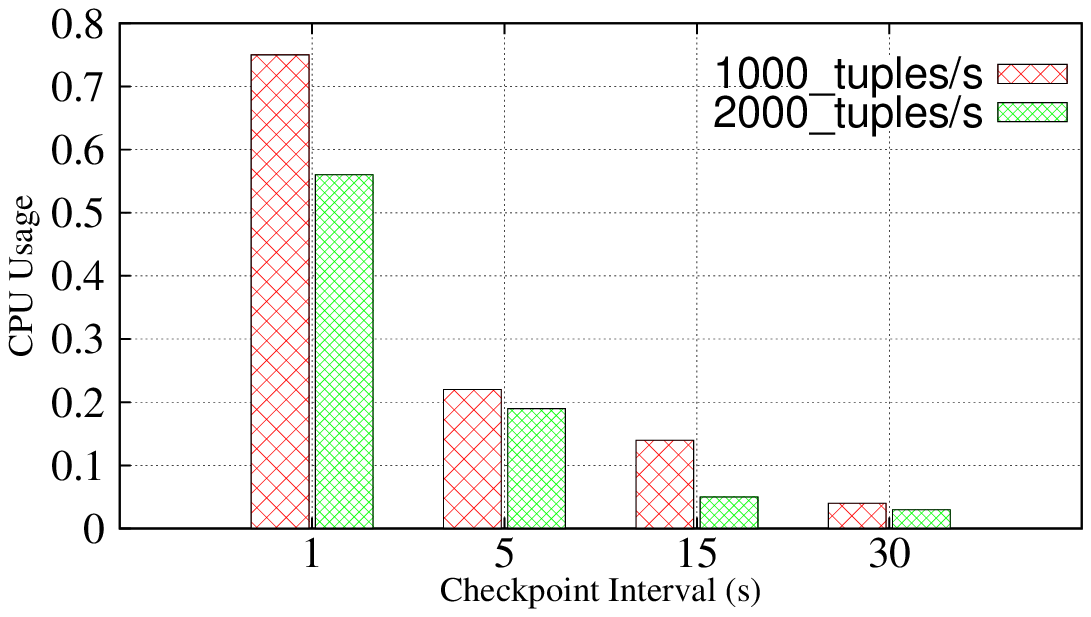}
                      \caption{Resource usage of maintaining checkpoints, window length: 30 seconds.}
                      \label{fig:resource}
      \end{minipage}
  \end{figure*}

In the first set of experiments, we study the recovery efficiencies of different
fault-tolerance techniques, including checkpoint, which is used
in Spark Streaming,  source replay, which is the default fault-tolerance technique in 
Storm, and active replication.  In Storm,
if failure happens, the source data will be reprocessed from scratch through the
whole topology to rebuild the states of the tasks. 

We implement a topology that consists of 1 source operator and 4 synthetic
operators. The structure of this topology is depicted in Figure~\ref{fig:stopo}.
The source operator consists of totally 16 tasks, which are on average deployed
on 4 nodes.  All of the source tasks produce input tuples for their downstream
neighboring tasks in a specified rate (1000 tuples/s or 2000 tuples/s).  The
degree of parallelization of operators $O_1$, $O_2$, $O_3$ and $O_4$ are set as
8, 4, 2 and 1 respectively. Each task in  $O_1$ receives inputs from two source
tasks and each task in $O_2$, $O_3$ and $O_4$ receives inputs from two upstream
neighboring tasks. The primary replicas of the 15 synthetic tasks are evenly
distributed among the 15 nodes. In addition, there are another 15 nodes used as
the backup nodes to store the checkpoints and to run the active replicas.

Each of the four synthetic operators maintains a sliding window whose sliding 
step is set as 1 second and window interval varies from 10 seconds to 30 seconds.
The state of each task of a synthetic operator is composed by the input data 
within the current window interval. The largest state size of a task is equal to
the result of the input rate multiplies the window interval. The selectivity of
the synthetic operator is set as $0.5$.

\textit{\textbf{Single Node Failure.}} 
Figure~\ref{fig:single} presents the recovery latencies of single node failures with 
various input rates and window intervals using different fault-tolerance techniques. 
For active replication, we vary the intervals of trimming the output buffer of
a task replica, which is equivalent to the frequency of synchronizing the
replica with its primary task. One can see that the active approach has much
lower recovery latency than the passive approaches and the changes of window
intervals and input rates have little influence. On the other hand, the recovery
latencies with both Checkpoint and Storm increase proportionally with the input
rate, as a higher input rate results in more tuples to be replayed
during recovery for both approaches. Furthermore, the recovery latency with
Checkpoint increases with the checkpoint interval. This is because the number of tuples
that need be reprocessed to recover the task state will increase
with the checkpoint interval. 

As Storm will have to replay more source data with longer window intervals,
one can see that the recovery latency of Storm with 30-second windows is
higher than those with 10-second windows.  Another factor that influences the
recovery latency of Storm is the location of the failed task in the topology,
because the replayed tuples will be processed by all the tasks located between
the tasks of the source operator and the failed tasks. Thus the recovery latency of
Storm is higher than that of Checkpoint in most of the cases in
this experiment. Here, we record the recovery latencies of tasks in different
locations within the topology in Storm and report their average values.

\textit{\textbf{Correlated Failure.}} We inject a correlated failure by killing 
all the nodes on which the primary replicas of the tasks are deployed.
In Figure~\ref{fig:correlated}, one can see that active replication has much lower recovery latency than 
Checkpoint and Storm. Furthermore,  active replication with a shorter 
synchronization period leads to faster failure recovery. This is because, with 
a longer synchronization period, an active replica will send more buffered tuples 
to its downstream tasks if its primary is failed. 
On the other hand, the recovery latency of Checkpoint increases rapidly 
with the increase of input rate and  checkpoint interval.  
Storm has a lower recovery latency than that of Checkpoint with a 
30-second checkpoint interval.  This is because the window intervals in this 
set of experiments are relatively short. In Storm, to build the window states, all the sources 
tuples belonging to the unfinished window instances in the failed tasks
 will be replayed, whose number increases linearly with the window length. 
 While for the recovery with Checkpoint, the number of  tuples that should be 
 reprocessed to recover a failed task is at most equal to the value 
 of the input rate multiplies the checkpoint interval.

By comparing the experimental results presented in Figure~\ref{fig:single} and 
Figure~\ref{fig:correlated}, it can be seen that the recovery latency with active replication 
is lower than the passive approaches and is relatively 
stable under the scenarios of various input rates and window intervals. 
Moreover, the benefits of using active replication are larger in the case of
correlated failure than in the case of single node failure.  This is
because 
some synchronization operations will be performed during the
recovery of correlated failures.  

The latency of failure recovery with checkpoint can be reduced by setting a short checkpoint
interval.  However, the resource usage of maintaining checkpoints varies with
different checkpoint intervals. Figure~\ref{fig:resource} presents the ratio of
the CPU usage of maintaining checkpoint to that of normal computation within a
task. We can see that  the CPU usage of maintaining checkpoints increases
quickly with shorter checkpoint intervals and making checkpoint with very short
intervals such as one second is prohibitively expensive. Although active replication
consumes more recourses than the passive approach, the low-latency recovery of active 
replication makes it meaningful in the context of MPSPEs.

\begin{figure}[htb]
    \centering
    \subfigure[rate:1000 tuples / sec]{
        \label{fig:h1}
        \centering
        \includegraphics[width=0.45\linewidth]{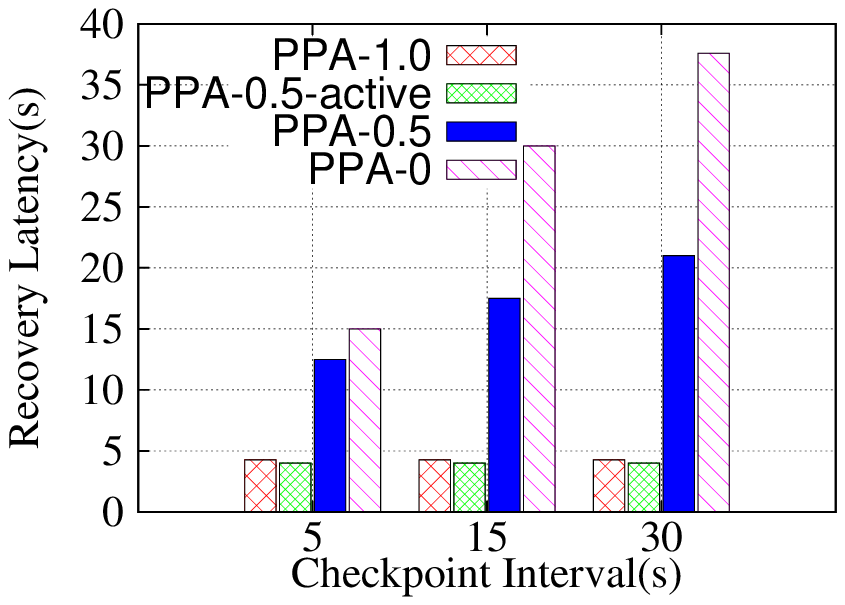}
    }
    \hspace{-10pt}
    \subfigure[rate:2000 tuples / sec]{
        \label{fig:h2}
        \centering
        \includegraphics[width=0.45\linewidth]{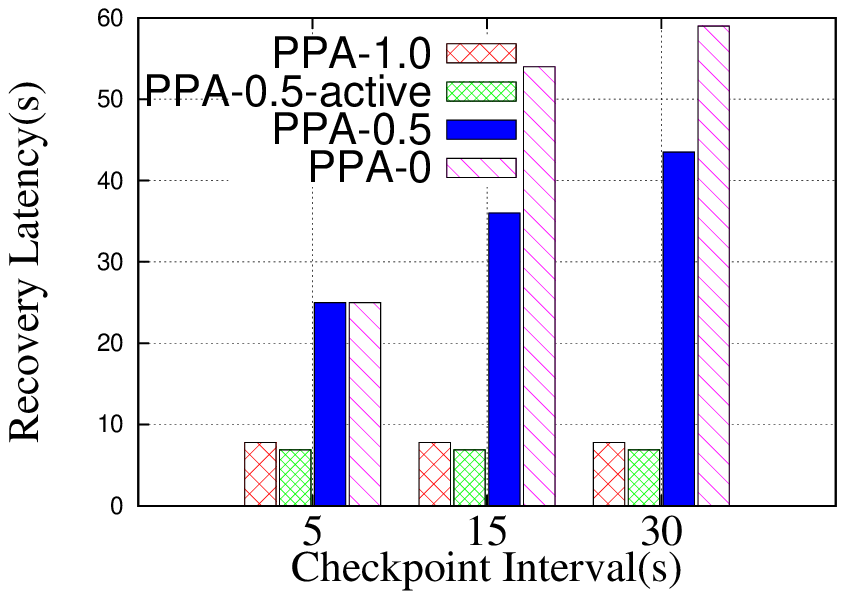}
    }
    \caption{Recovery latency  of a correlated failure with PPA, window length : 30 seconds.
    PPA-0.5-active indicates the recovery latency of
    actively replicated tasks in plan PPA-0.5.}\label{fig:hybrid}
\vspace{-5pt}
\end{figure}

\textit{\textbf{Recovery with PPA.}} We conducted experiments to study the
performance of PPA with three active replication plans denoted as
PPA-1.0, PPA-0.5 and PPA-0 respectively.  These PPA plans consume various amount
of resources for active replication.  In PPA-1.0, all the tasks in the topology
will be actively replicated.  PPA-0.5 is a hybrid replication plan where only
half of the tasks have active replica.  PPA-0  is a purely
passive replication plan where all the tasks are only replicated with
checkpoint. The results are presented in Figure~\ref{fig:hybrid}.
As the failed tasks with active
replicas will be recovered faster than those using checkpoints, the overall
recovery latency of PPA-0.5 is higher than that of PPA-1.0 but lower than that
of PPA-0. Note that with PPA-0.5, the recovery latencies of tasks with active replicas
(denoted as PPA-0.5-active in Figure~\ref{fig:hybrid})
are much lower than that of recovering all the failed
tasks (denoted as PPA-0.5 in Figure~\ref{fig:hybrid}). The recoveries of 
PPA-0.5-active consume slightly less time than PPA-1.0, this is because the number of
actively replicated tasks recovered in PPA-0.5-active is only the half of that in PPA-1.0. 
This set of experiments illustrate that the purely active replication plan outperforms the 
hybrid and purely passive plan regarding the recovery latency. With a hybrid plan, 
as the recoveries of actively replicated tasks finish earlier than that of the passively replicated ones,
PPA can generate tentative outputs without waiting for the slow recoveries of passively replicated
tasks.

\vspace{-5pt}

\subsection{Tentative Output Quality}\label{sec:tentative}

\begin{figure}[h!]
 \centering
     \includegraphics[scale=0.6]{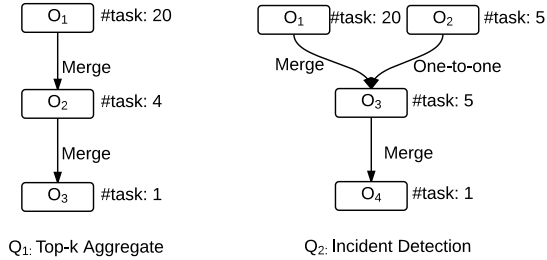}
     \caption{\emph{Top-k} aggregate query$(Q_1)$ and incident detection query$(Q_2)$
     in the scale of operator.}
    \label{fig:eirtopo}
\end{figure}

%
%



We implement two sliding window queries whose inputs are, respectively, from
real and synthetic datasets. For each query, we define an accuracy
function based on its semantic. 

$Q_1$ is a sliding-window query that calculates the \emph{top-100}
hottest entries of the official website of World Cup 1998. The input dataset
is the server access log during the entire day of June 30, 1998~\cite{data:wc98},
which consists of in total $73,291,868$ access records. In the experiments, 
we replay the raw input stream in a rate which is $48$ times faster
than the original data rate. We implement this query as a topology that conducts
hierarchical aggregates, which is a common computation in data stream
applications. The structure of this topology is depicted  in 
Figure~\ref{fig:eirtopo}.  Input tuples are partitioned to the tasks in $O_1$ 
by their server ids.  Tasks in $O_1$ split the input stream into a set of 
consecutive slices, each consisting of 100 tuples, and calculate their aggregate results. 
For every 100 input tuples, tasks in $O_2$ will conduct a
merge computation and send the results to the single task in $O_3$, which
periodically updates the globally \emph{top-100} entries for every 100 input
tuples.

$Q_2$ is a sliding-window query that detects the traffic incidents
resulting in traffic jams. The window interval is 5  minutes and the sliding
step is 10 seconds. As relevant datasets for this query are not publicly
available due to privacy considerations, we generate a synthetic dataset in a
community-based navigation application. There are two streams in this dataset:
the user-location stream and the incident stream. The rate of the user-location
stream is set as 20,000 location records per second. The incident stream is
composed of user-reported incident events and the time interval between two consecutive
incidents is set as 2 seconds. We distribute 100,000 users among 1000 virtual
road segments following the Zipfian distribution (with parameter $s=0.5$).  The
incident probability of a segment is set to be proportional to the number of
users located on it. If an incident occurs on a segment, all the users on this
segment will report an incident event. The topology of $Q_2$ is presented 
in Figure~\ref{fig:eirtopo}. Tasks in $O_1$ receive the user-location records and
calculate the average speed of each segment per second.  Tasks in $O_2$ combine 
the user-reported incident events into distinct incident events.  $O_3$ joins the
segment-speed stream from $O_1$ and the distinct-incident stream from $O_2$. The
outputs of tasks in $O_3$ are the incidents that incur traffic jams. $O_4$
aggregates the outputs of $O_3$. 

  \begin{figure}[h!]
      \subfigure[Query: Q1.]{
          \label{fig:wc98_ofic}
          \centering
          \includegraphics[width=0.48\linewidth]{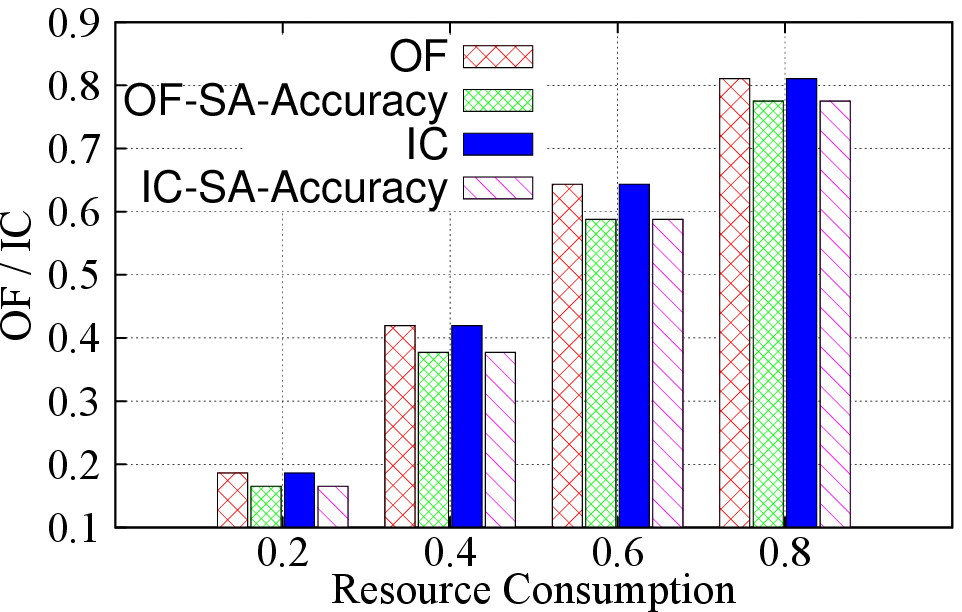}
      }
      \hspace{-10pt}
      \subfigure[Query: Q2.]{
          \label{fig:incident_ofic}
          \centering
          \includegraphics[width=0.48\linewidth]{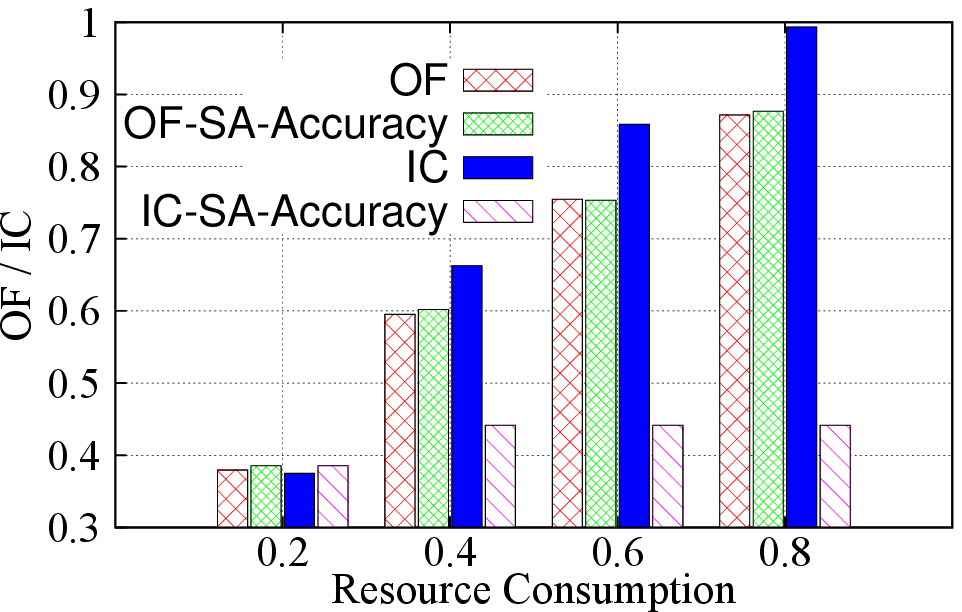}
      }
      \caption{Comparing the values of OF/IC and the query accuracy.  OF-SA-Accuracy
      (or IC-SA-Accuracy) denotes the actual query accuracies of the PPA plans generated 
      using the structure-aware(SA) algorithm with OF (or IC) as the optimization metric. 
           }\label{fig:ofic}
  \end{figure}
  
    \begin{figure}[h!]
        \subfigure[Query: Q1.]{
                \label{fig:wc98_algorithm}
                \centering
                \includegraphics[width=0.48\linewidth]{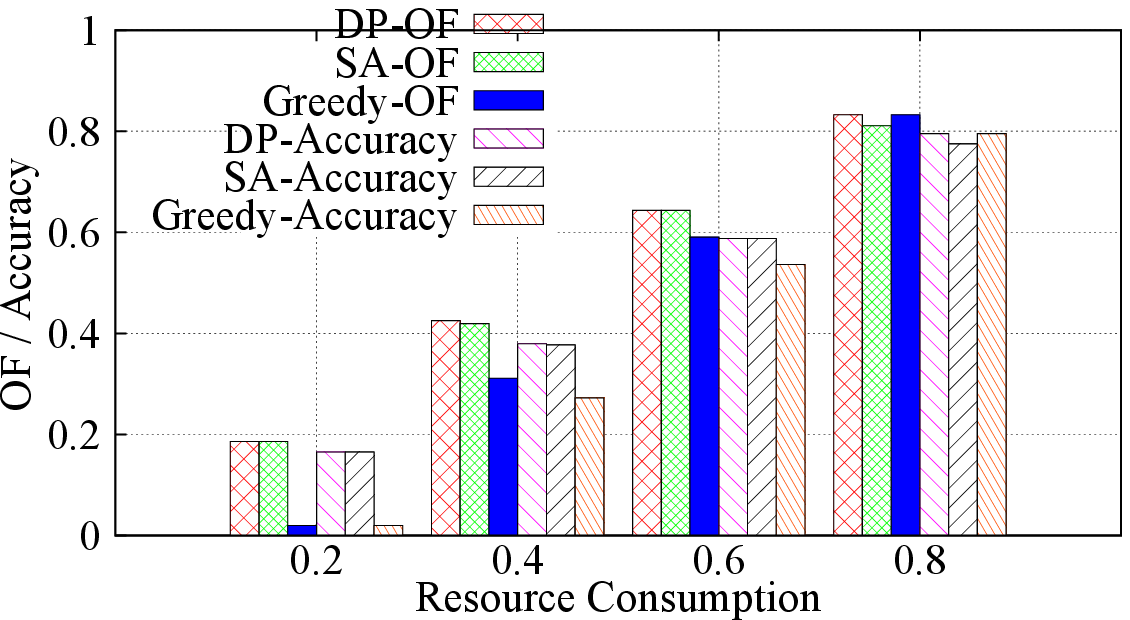}
            }
           \hspace{-10pt}
        \subfigure[Query: Q2.]{
                \label{fig:incident_algorithm_ic}
                \centering
                \includegraphics[width=0.48\linewidth]{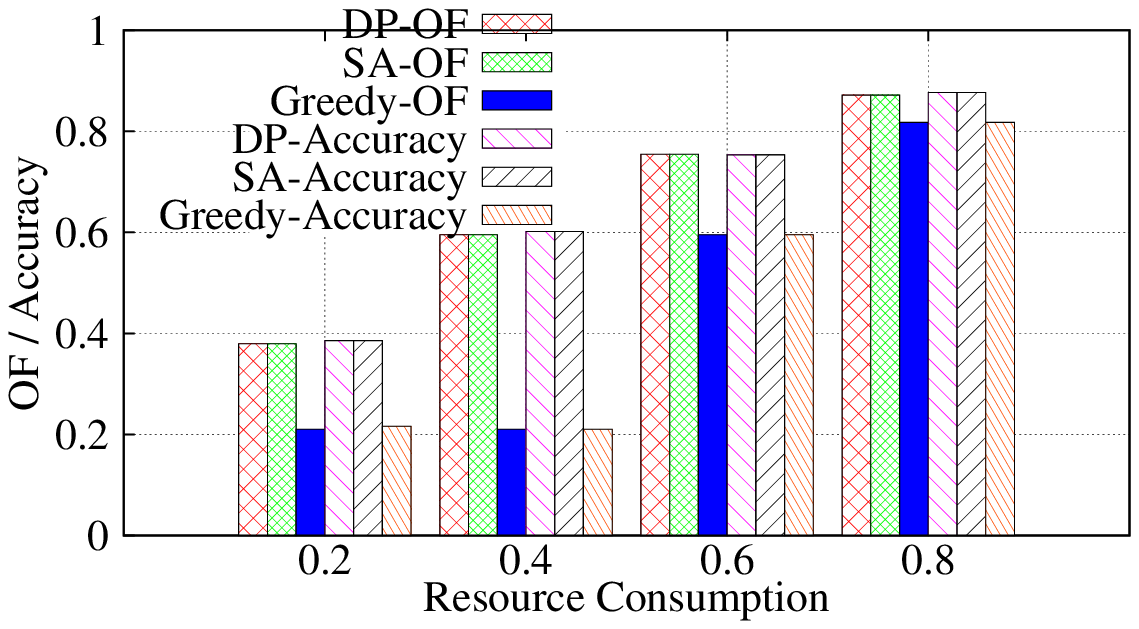}
            }
        \caption{Comparing the values of OF and the actual query accuracies of the PPA plans
        which are generated by the dynamic programing algorithm(DP), structure-aware algorithm(SA) 
        and  greedy algorithm(Greedy) respectively.}\label{fig:algorithm}
    \end{figure}

 \begin{figure*}[htb]
  \vspace{-10pt}
     \centering
     \subfigure[Workload skewness]{
         \label{fig:skew}
         \centering
         \includegraphics[width=0.24\linewidth]{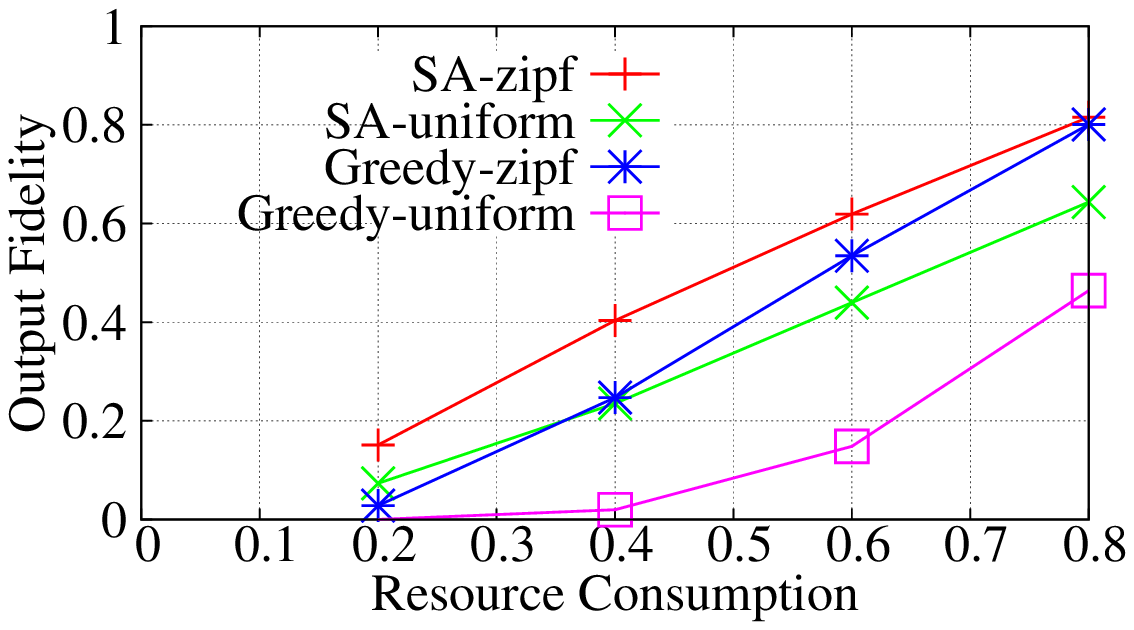}
     }
     \hspace{-10pt}
     \subfigure[Degree of parallelization]{
         \label{fig:para}
         \centering
         \includegraphics[width=0.24\linewidth]{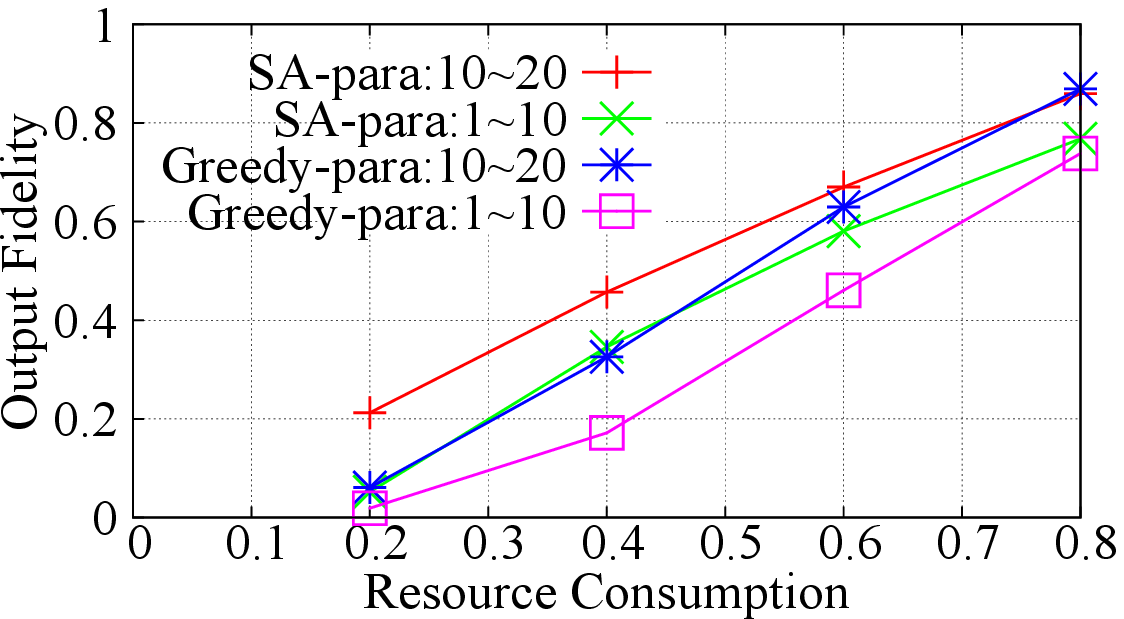}
     }
     \hspace{-10pt}
     \subfigure[Full partitioning]{
             \label{fig:nofull}
             \centering
             \includegraphics[width=0.24\linewidth]{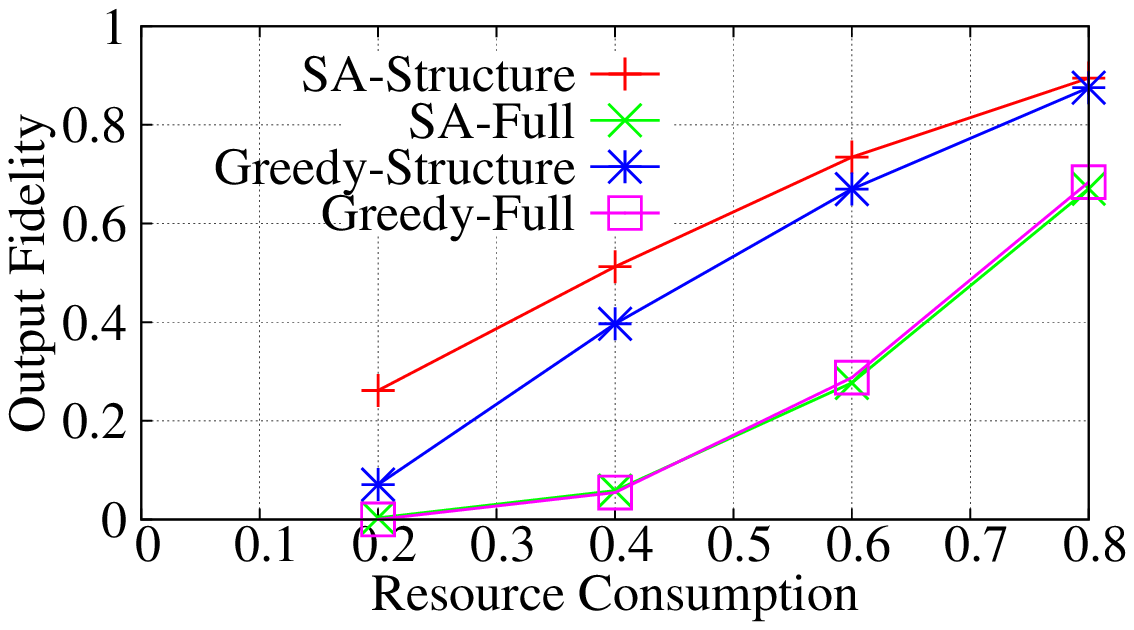}
         }
         \hspace{-10pt}
     \subfigure[Fraction of join operators]{
             \label{fig:join}
             \centering
             \includegraphics[width=0.24\linewidth]{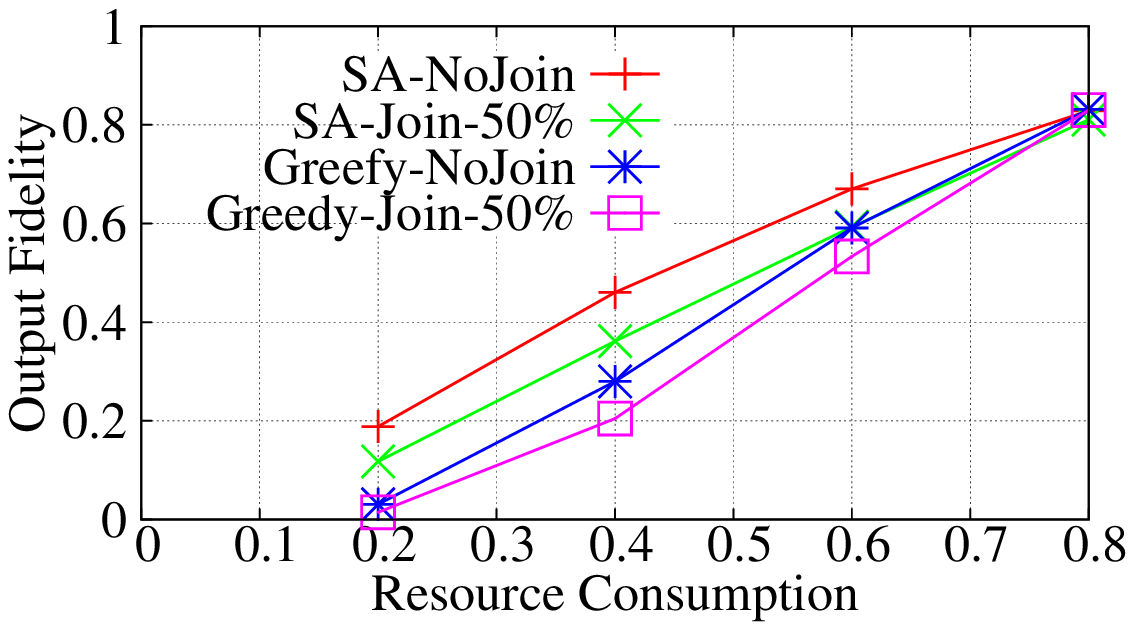}
         }
     \caption{Comparing OF of SA and Greedy algorithm with random topologies
     of various specifications, number of operators is set as a random integer between $5$ and $10$. 
     (a): The workloads of tasks within an operator
     are distributed in uniform or Zipfian distribution (with parameter $s = 0.1$). (b): The
     degree of operator parallelization is a random number between different ranges. (c): 
     Topologies are either structured topology or full topology. (d):  The fraction of join 
     operators in the topologies is set as $0$ or $50\%$.}\label{fig:handmade}
 \end{figure*}

\textbf{\textit{Validation of the OF metric.}} 
In this set of experiments, we examine whether OF can 
predict the actual quality of the tentative output. We compare it with the 
Internal Completeness $\left( IC \right)$ metric proposed in~\cite{Paolo:2014},
which measures the fraction of the tuples that are expected to be processed by
all the tasks in case of failures compared to the case without failures. A
fundamental difference between OF and IC is that, OF takes the correlations of
task's input streams into account.

By denoting the tentative outputs as $S_T$ and the accurate outputs of $Q_1$ as
$S_A$, we define the query accuracy of $Q_1$ as: $\frac{|S_T \bigcap
S_A|}{|S_A|}$.  Figure~\ref{fig:wc98_ofic}   shows the OF (or IC) values and the 
actual query accuracies of the PPA plans generated using  the OF (or IC) metric.  
The results show that both OF and IC provide good predictions of the accuracy of
typical top-k queries. This is because both OF and IC provide accurate
estimations of the completeness of the inputs for aggregate queries, such as
top-k, and such queries' output accuracies highly depend on the completeness of
their inputs.
The accuracy function of $Q_2$ is defined as $\frac{|I_T \bigcap
I_A|}{|I_A|}$, where $I_T$ is the set of tentative incidents generated with
correlated failure and $I_A$ is the set of accurate incidents generated without
failure. As shown in Figure~\ref{fig:incident_ofic}, the accuracy values 
are generally quite close to the values of OF. 
On the other hand, with more available resources, we can generate PPA
plans with higher IC values. However, such plans do not have higher query
accuracies. This is because IC fails to consider the correlation of tasks' input
streams and hence cannot provide a good accuracy prediction for queries with
joins. This result clearly indicates the importance of distinguishing join
operators in predicting output accuracies. 

\textbf{\textit{Comparing Various Algorithms.}} In this set of experiments, we generate
PPA plans for $Q_1$ and $Q_2$ using the dynamic programing algorithm(DP), 
the structure-aware algorithm(SA) and the greedy algorithm respectively and compare
their performances. Results presented in Figure~\ref{fig:algorithm} show that SA is
quite close to DP, which generates the optimal PPA plan, 
in both OF and the actual query accuracy.  Greedy has the worst
performance in the results of both queries. This is because Greedy fails to consider
that only complete MC-trees can contribute to the query outputs.

\subsection{Random synthetic topology}
To conduct a comprehensive performance study of PPA algorithms with various types of topologies, 
we implemented a random topology generator which can generate topologies with 
different specifications.
In the experiments, for each set of topology specifications, we generate 100
synthetic topologies and use them as the inputs of the structure-aware algorithm 
and the greedy algorithm to compare their performances in terms of OF. 
Due to the prohibitive complexity of the dynamic programing algorithm, we cannot complete
it for this set of experiments within a reasonable time so we do not include it here.
Query accuracies are not compared in this set of experiments, 
as we cannot derive the actual output accuracies for these randomized synthetic topologies.

In Figure~\ref{fig:handmade}, one can see that, SA outperforms the greedy
algorithm in all the combinations of topology specifications and active
replication ratios.  With smaller replication ratio, there is a greater
difference between SA and the greedy algorithm. This is because the greedy
algorithm is agnostic to the structure of the query topologies, and with a
smaller replication ratio, there is smaller probability that the tasks selected by the
greedy algorithms can form complete MC-trees that can contribute to the final output.
    
Figure~\ref{fig:skew} depicts the effects of workload skewness of tasks within
the operators. We can see that SA has better performance for topologies that
have higher skewness of task workloads. This is because, as the skewness of
workloads increases, the skewness of MC-trees' contributions to the value
of OF also increases and SA, by prioritizing tasks that are in the MC-trees, 
achieves higher OF values. In Figure~\ref{fig:para}, we report the results with varying
parallelization degrees of an operator. One can
see that increasing the parallelization degrees will also increase
the value of OF, because a higher parallelization degree slightly
increases the skewness of the workloads of the tasks in this set of experiments. As shown in
Figure~\ref{fig:nofull}, the OF of structured topologies
are generally higher than the full topologies. This is because within an operator 
using Full partitioning, the failure of any task
will reduce the input of all the downstream tasks. For full topologies, the
structure-aware algorithm generates active replication plan in the similar approach 
as the greedy algorithm does, thus their performances are close in this set of experiments. 
Figure~\ref{fig:join} presents the results with various fractions of operators being join operators.
For the same topology, OF decreases with more operators set as joins.  This is because
the loss of one input stream of a join operator will result in parts of the
other (correlated) input streams being useless. 

\section{Related Work}

\textbf{Fault-tolerance in SPE.} 
Traditional fault-tolerance techniques for SPEs could be categorized as passive
~\cite{Hwang:2005,Upadhyaya:2011,Kwon2008,Madsen:2015,Madsen:2014} and active
approaches~\cite{Hwang:2005,Balazinska:2008,jh2008}. 
The technique of delta checkpoint~\cite{Hwang2007} is used to reduce the size of
checkpoints. The authors in~\cite{Gu:2009} proposed techniques to reduce the
checkpoint overhead by minimizing the sizes of queues between operators, which
are part of the checkpoints. \cite{Martin:2011} proposed to utilize
the idle period of the processing nodes for active replication. Such
optimizations are compatible to our PPA scheme and can be employed in our
system. 

Spark Streaming~\cite{Zaharia:2013} uses Resilient Distributed Dataset (RDD) to
store the states of processing tasks. In case of failure, RDDs can be restored
from checkpoints or rebuilt by performing operations that were used to build it
based on its lineage. In other words, it adopts both the checkpoint-based and
the replay-based approaches.

For other large-scale computing systems, such as
Map-Reduce~\cite{dean2008mapreduce}, the overall job
execution time is a critical metric. However, for MPSPEs, it is the end-to-end
latency of tuple processing that matters, which makes the low-latency failure
recovery an important feature in the context of MPSPEs.  To reduce recovery
latency, authors in~\cite{CastroFernandez:2013,Zaharia:2013} proposed to use
parallel recovery and/or integrating fault tolerance with scale-out operations.
In parallel recovery, multiple tasks can be launched to recover a failed task
and each of them is recovering a partition of the failed one to shorten the
process of passive recovery. However, with a correlated failure, a large number
of failed tasks need to be recovered simultaneously. Then the possibilities of
fast scaling out and the degrees of parallel recovery would be constrained. 

Hybrid fault-tolerance approaches are proposed in~\cite{Upadhyaya:2011, Heinze:2015}.
In \cite{Upadhyaya:2011}, the objective is to minimize the total cost by choosing a 
passive fault-tolerance strategy, including upstream buffering, local checkpoint and 
remote checkpoint,  for each operator. \cite{Heinze:2015} uses either
active replication or checkpoint as the fault-tolerance approach for an operator.
The optimization objective in \cite{Heinze:2015} is to minimize the total processing
cost while satisfying the user-specified threshold of recovery latency, where only 
independent failure is considered. The work in~\cite{Zhang:2010} considers task
overloading, referred to as ``transient'' failure, caused by
temporary workload spikes. Upon a transient failure of a task, its
active replica will be used to generate low-latency output. Different from these
approaches, the trade-off of our work is between resource consumption and result
accuracy with correlated failures.  

\textbf{Tentative Outputs.}
Borealis~\cite{Balazinska:2008} uses active replication for fault 
tolerance and allows users to trade result latency for
accuracy while the system is recovering from a failure. More
specifically, if a failed node has no alive replica, Borealis will
produce tentative outputs if the recovery cannot be finished within a
user-defined interval. PPA adopts a similar mechanism for generating
tentative outputs but explores more on optimizing the accuracy of tentative
results. Previous work~\cite{Paolo:2014} attempts to dynamically assign computation 
resources between primary computation and active replicas
to achieve trade-offs between system throughput
and fault-tolerance guarantee. Their accuracy model, IC,  does not consider
the correlation of processing tasks' inputs streams, which is shown to be
inadequate in our experiments.
The brute-force algorithm proposed in \cite{Paolo:2014} which has a high
complexity as our dynamic programing does. 


A fault injection-based approach is presented in \cite{Jacques-Silva:2011}
to evaluate the importance of the computation units to the output accuracy,
which only considers independent failures.  Zen~\cite{Zen:2008} optimizes
operator placement within clusters under a correlated failure model, which
specifies the probability that a subset of the nodes fail together.  The
objective is to maximize the accuracy of tentative outputs after failures. As
operator placement is orthogonal to the planning of active replications, their
techniques can also be employed as a supplement to PPA. 

\textbf{Failure in Clusters.} 
Previous studies found that failure rates vary among different clusters and
the number of failures is in general proportional to the size of the
cluster~\cite{Schroeder:2006}. Correlated failures do exist and their scopes
could be quite large~\cite{Heath:2002,Nath:2006}. Hence considering
correlated failure is inevitable for a MPSPE that supports low-latency and
nonstop computations.

\section{Conclusion}
\balance
In this paper we present a passive and partially active (PPA)
fault-tolerance scheme for MPSPEs. In PPA, passive checkpoints are
used to provide fault-tolerance for all the tasks, while active
replications are only applied to selective ones according
to the availability of resources. A partially active replication plan is
optimized to maximize the accuracy of tentative outputs during failure
recovery. The experimental results indicate that upon a correlated failure,
PPA can start producing tentative outputs up to 10 times faster than 
the completion of recovering all the failed tasks. Hence PPA is suitable for
applications that prefer tentative outputs with minimum delay. The
experiments also show that our structure-aware algorithms can achieve up to one
order of magnitude improvements on the qualities of tentative outputs in comparing
the greedy algorithm that is agnostic to query topology structures, especially
when there is limited resource available for active replications. Therefore, to
optimize PPA, it is critical to take advantage of the knowledge of the query
topology's structure.


\balance

\bibliographystyle{abbrv}
{\renewcommand\baselinestretch{0.9}\selectfont
\bibliography{paper}
\par}


\end{document}